\documentclass[]{llncs}

\usepackage{booktabs}
\usepackage{amssymb,amsmath}
\usepackage{url}
\usepackage{tikz,tikz-inet,pgf,pgfplots}
\pgfplotsset{compat=1.13}
\usepackage[caption=false]{subfig}
\usepackage[ruled]{algorithm2e}
\pgfcreateplotcyclelist{mylist}
	{black,mark=none,thick 
}

\usepackage{amssymb}

\usepackage[figuresright]{rotating}

\begin{document}
\mainmatter              
\title{Project Makespan Estimation: Computational Load of Interval and Point Estimates}
\titlerunning{Project Makespan Estimation}  
%
\author{Maurizio Naldi\inst{1} \and Marta Flamini\inst{2}
}
\authorrunning{Naldi and Flamini} 
%
\tocauthor{Maurizio Naldi and Marta Flamini}
\institute{University of Rome at Tor Vergata, Rome, Italy,\\
\email{maurizio.naldi@uniroma2.it}
\and
International Telematic University UNINETTUNO, Rome, Italy,\\
\email{m.flamini@uninettunouniversity.net}}

\maketitle 

\begin{abstract}
The estimation of project completion time is to be repeated several times in the project planning phase to reach the optimal tradeoff between time, cost, and quality. Estimation procedures provide either an interval or a point estimate. The computational load of several estimation procedures is reviewed. A multiple polynomial regression model is provided for  major interval estimation procedures and shows that the accuracy in the probability model for activities is the most influential factor. The computational time does not appear to be an impeding factor, though it is larger for MonteCarlo simulation, so that the computational time can be traded off in search of a simpler estimation procedure.  

\end{abstract}



\section{Introduction}
Estimating a project duration (its makespan) is a major task during the planning phase. Duration is one of the three quantities (time, cost, and quality) involved in the search for an optimal tradeoff when defining a project structure \cite{BABU1996320}. Though we wish to complete a project as early as possible, its duration is determined by the amount of resources that the company in charge of the project can employ in the process. The search for the optimal tradeoff requires therefore a repeated evaluation of the project duration, since the company has to explore a variety of options, relaxing or redefining some constraints on the resource it may use (e.g., by facing the payment of sanctions for delay, or hiring additional resources to advance activities, or shifting activities along the timeline to take advantage of the availability of some resources).

Since the duration of most activities within the project is uncertain in nature, the duration of the overall project is actually a random variable. Evaluating the project duration is therefore an estimation task, which may mean estimating either one or more statistical parameters (typically the mean and standard deviation) or the full probability distribution of the project makespan.

The need to repeatedly estimate the project makespan (to explore the options available and the tradeoff with cost and quality) calls for estimation procedures that can be performed in a reasonable timeframe and provide the needed accuracy. Recently, Hall has included the estimation of project makespan and the probability distribution of project completion times in his list of open research problems in project management, highlighting the complexity of current estimation procedures as the factor impeding their adoption by companies \cite{hall2016research}. 

In this paper, we wish to assess the computational load of the estimation procedure and see whether it represents an actual hurdle in the effective usage of project makespan estimation procedures within a company. After describing in detail the elements of the project makespan problem in Section \ref{sec:problem}, we provide the following contributions:
\begin{itemize}
    \item we report the computational load of the major estimation procedures (Sections \ref{sec:point} and \ref{sec:interval});
    \item for the interval estimates (which determine an upper and a lower bound of the project makespan) we provide a multiple polynomial regression approximation of the computational load (Section \ref{sec:interval});
    \item we assess the relevance of the factors contributing to the computational load and provide a ranking (Section \ref{sec:interval});
    \item we compute some statistics concerning real project to envisage the actual size of typical project makespan estimation tasks (Section \ref{sec:pratica}).
\end{itemize}

\section{The project makespan problem}
\label{sec:problem}
The estimation of the project makespan is carried out through a graph model of the project itself. In turn, the computational cost of the estimation is certainly a function of the project size. In this section we briefly describe the alternative options for the graph model, the metric to represent the project size, and the aim of the estimation.

 If we refer to a graph which is the symbolic structure typically adopted as a schematic representation of the project, the entities representing the size of the graph are its nodes and its arcs. We have basically two choices for the correspondence between the real project and the graph: the Activity-on-node (AoN) model, where nodes are activities and arrows (arcs) indicate the precedence relationships, and the Activity-on-arc (AoA) model, which has arcs for activities with nodes being the starting and ending points. The AoN is frequently employed in practical, non-optimization situations, while the AoA is used in optimization settings. In this context, we can consider the size of  project to be represented by the number of activities to be carried out during the project itself. Therefore, we  consider the number of nodes in the AoN case and the number of arcs in the AoA case as a measure of the project size (in other contexts, the investment required by the project is instead considered as a measure of its size \cite{Naldi-ems-2015,naldi2016maximin,NaldiHHI,Noictw2016}).
 
 As to the object of the estimate, the project duration is actually a random variable, since it depends on the duration of the activities making up the project. Defining the random variables representing the activities is itself a difficult task. The Beta distribution is employed in the PERT method, while the triangular and the normal model have been employed in MonteCarlo simulation. Defining the probability distribution of the activities is not an easy task \cite{williams1992practical}, but has a significant impact on the quality of the estimates, so that it has been suggested to heavily exploit the historical information regarding similar projects \cite{kirytopoulos2008PERT}. Alternative methods to describe the uncertainty surrounding the project activities rely on the use of fuzzy logic \cite{lorterapong1996project}.
 
In this paper we rely on the probabilistic approach, so that the task is estimating some parameters of the project makespan. In particular we classify the estimates (and the estimation methods) into two classes: interval estimates and point estimates. An interval estimate provides a range of values that contains the true project duration. This boils down to providing two figures: a lower bound and an upper bound. If the probability distribution of activities has a finite support, the bounds are tight, i.e. the project is guaranteed to end not before a certain date and not later than another date. Otherwise, the bounds are themselves defined in a probability framework, i.e. the probability that the project ends between the two bounds is a (typically large) value different from 1. Point estimates instead provide a single figure for the project duration, typically the mean or the median value.

\section{Makespan Interval Estimates}
\label{sec:interval}
We consider first the class of interval estimates.
We refer to the selection of methods considered in \cite{ludwig2001computational}:
\begin{itemize}
    \item Upper and lower bound proposed by Kleindorfer
    \item Upper bound by Dodin
    \item Upper and lower bound by Spelde
\end{itemize}

In all cases, the authors employ an AoA network model denoted $N(V,A)$, where nodes correspond to project events and arcs correspond to project activities.  $F_{[v]}$ and $F_a$ are the distribution functions associated respectively to a node (event) $v \in V$ and to an arc (activity) $a \in A$.

In \cite{kleindorfer1971bounding} Kleindorfer provides both a stochastic upper and a stochastic lower bound on the makespan distribution. For each node, Kleindorfer derives the bounds on the probability distribution of the event starting time (when all the activities leading to that event have been completed), considering either their independence or their correlation. The upper bound (applying in the case of independence of activities) for $F_{[v]}$ is the \emph{product}, over all the nodes preceding $v$, of the convolution between $F_{[p(v)]}$ and $F_a$ where $p(v)$ is a node preceding $v$ in the network and $a$ is the activity represented by the arc $a = (p(v),v)$:

\begin{equation}
\begin{centering}
F_{[v]} = \Pi_{p(v)\in N, a = (p(v),v)}  F_{[p]} \ast F_a
\end{centering}
\end{equation}

The upper bound by Kleindorfer is exact for node $v$ if the all paths from the source node $s$ of the network to $v$ are disjoint.
The lower bound for $F_{[v]}$ (accounting for the case of correlation) is the \emph{minimum} among all the convolution between $F[p(v)]$ and $F_a$.

\begin{equation}
\begin{centering}
F_{[v]} = \min_{p(v)\in N,  a = (p(v),v)}  F_{[p]} \ast F_a
\end{centering}
\end{equation}

The complexity of the algorithm is $\mathcal{O}(\vert A \vert · (\textrm{conv}(sp)+\textrm{prod}(sp)))$ where prod(sp)
and conv(sp) denote the complexity of the product and convolution operation depending
on the number of supporting points $sp$ employed to describe the probability distribution function.



In \cite{dodin1985bounding} Dodin instead provides an upper bound on the makespan distribution by reducing the network applying the standard series-parallel reduction. If there exists a node that has one ingoing (outgoing) arc and more than one outgoing (ingoing) arcs, standard series-parallel reduction are not feasible, hence duplication is performed. 
Such operation transforms the network such that new series-parallel reduction are allowed. When a series reduction is performed, a set of arcs (activities) $a_{i_1} \ldots a_{i_n}$ is reduced to a single arc (activity) $h$ and $F_h$ is the \emph{convolution} of all the $F_{a_{i_1}} \ldots F_{a_{i_n}}$. When a parallel reduction is performed a set of parallel arcs $a_{i_1} \ldots a_{i_n}$ is reduced to a single arc $h$, and $F_h$ is the \emph{product} of all the distribution function $F_{a_{i_1}} \ldots F_{a_{i_n}}$. 
When a duplication of arc $h$ is performed a set of arcs $a_{i_1} \ldots a_{i_n}$ are added to the network and $F_{a_{i_1}} \ldots F_{a_{i_n}}$ are all equal to $F_h$.
After applying the reduction operations, the network is reduced to a single arc and the distribution function of such arc is the result of all the operations described above.

The complexity of Dodin's algorithm is $\mathcal{O}(\vert A \vert (\textrm{conv}(sp)+\textrm{prod}(sp)) + \vert A \vert \textrm{Max\_out\_degree})$, where $\textrm{Max\_out\_degree}$ is the maximum out-degree of the nodes of the network.

Finally, Spelde provides both a stochastic upper bound and a stochastic lower bound on the makespan distribution \cite{spelde1976stochastische}. The lower bound is obtained considering a set of pairwise disjoint paths (not necessarily complete $(s,t)$-paths, also partial paths) and is given by the maximum length of such path. The length of a path is computed by considering the sum of the expectations of the processing times of the activities of the path. The complexity of the algorithm is $\mathcal{O}(\vert A \vert · (\textrm{conv}(sp) + \textrm{prod}(sp)) + \vert A \vert^{2})$. 

Analogously, for the upper bound Spelde calculates all the $(s,t)$-paths and the maximum stochastic length represents the upper bound for the makespan distribution.

For the four methods so far described, Ludwig et alii conducted some experiments on a Sun Ultra 1 with 143 MHz clock pulse operating under Solaris 2.6 with 64 MB of memory, using code written in C++ \cite{ludwig2001computational}. They provided the CPU time needed to estimate the bounds on the makespan for a selection of network sizes (embodied by the number of activities) and supporting points (which provide the accuracy of the representation of the probability distribution of activity duration). Even for a project made of 1200 activities, the average CPU time is below 2 minutes for all the methods. 

However, since their results are reported as a table, they do not provide a functional relationship among those variable, which could help understand the shape of the trend and the impact of each variable (though the figures refer to the computational environment employed by Ludwig et alii in \cite{ludwig2001computational}, the  relationship should just scale with a different computational power). Here we propose a multiple polynomial regression as a model to describe that relationship. The average CPU time $Y$ is represented as a second degree polynomial of the number of activities $X_{1}$ making up the project and the number $X_{2}$ of supporting points:
\begin{equation}
\label{regex}
    Y = a_{00} + a_{10}X_{1} + a_{01}X_{2} + a_{11}X_{1}X_{2} + a_{20}X_{1}^{2} + a_{02}X_{2}^{2}. 
\end{equation}
After a numerical regression on the data provided in \cite{ludwig2001computational} for the average CPU time, we obtain the coefficient values shown in Table \ref{tab:polyreg}.

\begin{table}[]
    \centering
    \begin{tabular}{lcccccc}
    \toprule
         & $a_{00}$ & $a_{10}$ & $a_{01}$ & $a_{11}$ & $a_{20}$ & $a_{02}$\\
         \midrule
       Kleindorfer (upper)  & $22.\overline{3}$ & -0.0308 & -0.4476 & $5.96\cdot 10^{-4}$ & $1.\overline{1}\cdot 10^{-6}$ & $1.71\overline{6}\cdot 10^{-3}$\\
       Kleindorfer (lower) & 20.883 & $-2.818\cdot 10^{-2}$ & -0.4333 & $5.897\cdot 10^{-4}$ & $-1.852\cdot 10^{-7}$ & $1.67\overline{3}\cdot 10^{-3}$\\
       Dodin & $21.8\overline{3}$ & $-2.92\cdot 10^{-2}$ & -0.4465 & $5.92\overline{6}\cdot 10^{-4}$ & $4.63\cdot 10^{-7}$ & $1.64\cdot 10^{-3}$\\
       Spelde & $0.0958\overline{3}$ & $-2.2\cdot 10^{-4}$ & $-1.857\cdot 10^{-3}$ & $1.148\cdot 10^{-5}$ & $9.259\cdot 10^{-8}$ & $8.\overline{3}\cdot 10^{-6}$\\
       \bottomrule\\
    \end{tabular}
    \caption{Polynomial regression coefficients for the CPU time}
    \label{tab:polyreg}
\end{table}

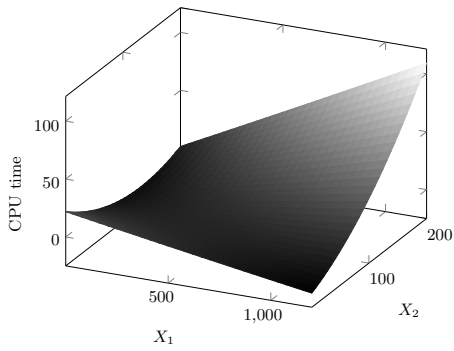
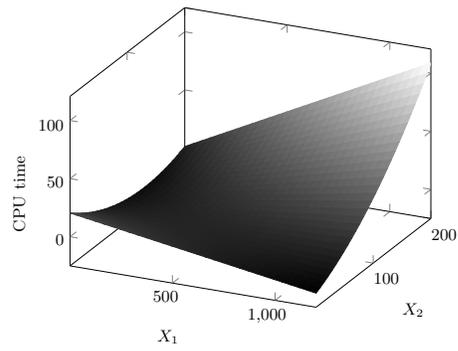
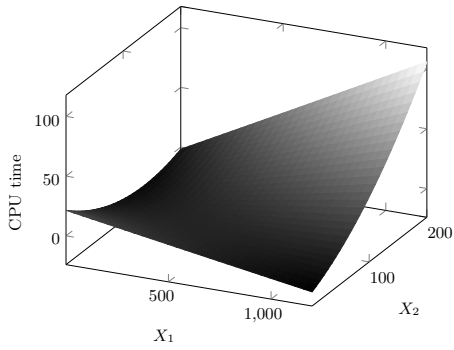
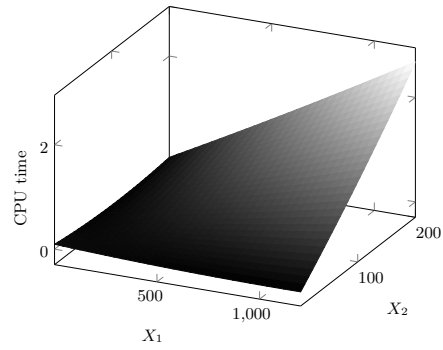
\begin{figure}
    \centering
\subfloat[Kleindorfer (upper)]{%
\begin{tikzpicture}[scale=0.7]
	\begin{axis}[xlabel=$X_{1}$,ylabel=$X_{2}$,zlabel=CPU time,colormap/blackwhite]
	\addplot3[
		surf,
		shader=flat,
		samples=30, domain=1:1200,y domain=1:200]
	{1.111E-6*(x^2)+5.96E-4*x*y+1.716666E-3*(y^2)-0.0308*x-0.4476*y+22.333};
	\end{axis}
\end{tikzpicture}
}
\subfloat[Kleindorfer (lower)]{
\begin{tikzpicture}[scale=0.7]
	\begin{axis}[xlabel=$X_{1}$,ylabel=$X_{2}$,zlabel=CPU time,colormap/blackwhite]
	\addplot3[
		surf,
		shader=flat,
		samples=30, domain=1:1200,y domain=1:200]
	{-1.852E-7*(x^2)+5.897E-4*x*y+1.673333E-3*(y^2)-0.0281888*x-0.4333*y+20.883};
	\end{axis}
\end{tikzpicture}
}

\subfloat[Dodin]{
\begin{tikzpicture}[scale=0.7]
	\begin{axis}[xlabel=$X_{1}$,ylabel=$X_{2}$,zlabel=CPU time,colormap/blackwhite]
	\addplot3[
		surf,
		shader=flat,
		samples=30, domain=1:1200,y domain=1:200]
	{4.63E-7*(x^2)+5.92666E-4*x*y+0.00164*(y^2)-0.02923*x-0.4465*y+21.8333};
	\end{axis}
\end{tikzpicture}
}
\subfloat[Spelde]{
\begin{tikzpicture}[scale=0.7]
	\begin{axis}[xlabel=$X_{1}$,ylabel=$X_{2}$,zlabel=CPU time,colormap/blackwhite]
	\addplot3[
		surf,
		shader=flat,
		samples=30, domain=1:1200,y domain=1:200]
	{9.259E-8*(x^2)+1.148E-5*x*y+8.333E-6*(y^2)-2.2222E-4*x-1.857E-3*y+0.0958333};
	\end{axis}
\end{tikzpicture}
}
\caption{CPU time for bound estimation}
\label{fig:boundcpu}
\end{figure}

The behaviour resulting from the polynomial regression is shown in \figurename~\ref{fig:boundcpu}, which provides a view of how the two components contribute to the CPU time. Though the overall fit is quite good (as shown by the R-squared which is 0.997 for Kleindorfer and Dodin, and an even larger 0.999 for Spelde), it must be kept in mind that their relative error may be large for lowest values of supporting points (where, however, computation times are so small as to make their correct estimation irrelevant), as shown in Table \ref{tab:error}.

\begin{table}[]
    \centering
    \begin{tabular}{cccccc}
    \toprule
    Activities & Supp. Points & Kleindorfer (upper) &	Kleindorfer (lower)	& Dodin	& Spelde\\
    \midrule
    300	 & 	50	 & 	-57.98	 & 	-54.96	 & 	-152.00	 & 	-27.34	 \\
300	 & 	100	 & 	95.32	 & 	91.09	 & 	55.92	 & 	7.37	 \\
300	 & 	200	 & 	-3.57	 & 	-3.81	 & 	-4.50	 & 	1.75	 \\
600	 & 	50	 & 	-16.02	 & 	-17.35	 & 	-22.27	 & 	11.90	 \\
600	 & 	100	 & 	9.44	 & 	7.16	 & 	8.60	 & 	3.09	 \\
600	 & 	200	 & 	-0.79	 & 	-0.63	 & 	-0.97	 & 	-2.61	 \\
900	 & 	50	 & 	19.96	 & 	15.40	 & 	14.25	 & 	-27.75	 \\
900	 & 	100	 & 	-4.97	 & 	-4.31	 & 	-5.62	 & 	-33.43	 \\
900	 & 	200	 & 	0.29	 & 	0.50	 & 	0.49	 & 	35.10	 \\
1200	 & 	50	 & 	46.10	 & 	44.93	 & 	31.50	 & 	3.62	 \\
1200	 & 	100	 & 	-11.06	 & 	-10.21	 & 	-12.75	 & 	-3.01	 \\
1200	 & 	200	 & 	1.19	 & 	0.92	 & 	1.03	 & 	0.80	 \\  
\bottomrule\\
    \end{tabular}
    \caption{Relative error [\%] of polynomial regression}
    \label{tab:error}
\end{table}

The relative importance of the terms involved in the regression can be assessed through the correlation matrix, following the approach described, e.g., in \cite{cohen2013applied}. Since the independent variables (number of activities and supporting points) are not measured but chosen when planning the simulation task, their correlation is not relevant, and we can limit ourselves to compute the correlation between the CPU time (Y) and each regressor involved in the regression analysis, i.e., the five terms $X_{1}$, $X_{2}$, $X_{1}\cdot X_{2}$, $X_{1}^{2}$, and $X_{2}^{2}$ as introduced in Equation (\ref{regex}). For all methods we see that the regressor more closely following the CPU time is the linear mixed term $X_{1}\cdot X_{2}$, i.e. the product of the number of activities and the number of supporting points, whose correlation is very close to 1. Though lower (roughly 0.8), the relevance of the number of supporting points (and their square) is however much larger than that of the number of activities.

\begin{table}[htbp]
    \centering
    \begin{tabular}{lccccc}
    \toprule
        Method & $X_{1}$ & $X_{2}$ & $X_{1}\cdot X_{2}$ & $X_{1}^{2}$ & $X_{2}^{2}$\\
        \midrule
        Kleindorfer (upper) &  0.406 &	0.824 &	0.965 &	0.400 &	0.832 \\
        Kleindorfer (lower) & 0.407 &	0.825 &	0.966 &	0.400 &	0.833\\
        Dodin & 0.422 &	0.811 &	0.969 &	0.416 &	0.819\\
        Spelde & 0.507 & 0.760 &	0.966 &	0.495 &	0.764\\
        \bottomrule\\
    \end{tabular}
    \caption{Correlation between the CPU time and the regression regressors}
    \label{tab:correl}
\end{table}

\section{Makespan Point Estimates}
\label{sec:point}
An alternative to bounding the project completion time is represented by its point estimation, for which we consider the two major estimation approaches: PERT and MonteCarlo.

\subsection{PERT}

As to PERT, Hagstrom proved that computing the expected completion time of the project, or any point of the cumulative distribution function of the completion time (such as a quantile or the median value) in a PERT network, cannot be done in polynomial time, unless P=NP \cite{hagstrom1988computational}. 

However, for the problem of estimating the distribution function of the completion time, Yao and Chu proposed an approximation scheme, comparing it with a standard PERT or MonteCarlo simulation \cite{yao2007new}. Their approximation relies on a simplification of the Discrete Re-sampling Technique introduction by Dodin in \cite{dodin1985approximating}, where the actual pdf is replaced by a discrete version with a small number of sampling points. They considered 20 instances of a 100 node network with 135 activities (adopting the AoA paradigm) and activity durations following the normal, exponential or uniform model. The resulting estimation times were incredibly large for MonteCarlo simulation (2103 seconds) and very low for PERT (0.12 seconds), with their method being low though a bit larger than PERT (0.76 seconds). However, when factoring in the accuracy in the estimation of the mean, the PERT turned out to be quite inaccurate (its average error was 24\% versus the 2.42\% of their method), though being the fastest of the group. 

\subsection{MonteCarlo}

The straightforward approach to MonteCarlo simulation to estimate the project makespan is to generate a (large) number of instances of the graph describing the project, where each instance involves generating a duration for each activity in the project (drawn from a suitable probability distribution), solving the precedence relations, and computing the duration of the critical path. The computation time is then the product of the number of instances and the time needed for each instance (which can be considered in turn to be proportional to the number of activities in the project, whose durations we have to simulate). As to the number of instances, in his seminal paper \cite{vanSlyke}, van Slyke has provided some indicative figures for the quantities of interest, namely the expected makespan and its probability distribution, the variance of the project duration, and the probability of an activity lying on the critical path (the values have been later recalled in \cite{milian2008monte}). The reported number of instances needed to get the expected makespan with a 95\% accuracy was 10,000. The time needed to get the estimate on the (now vintage) IBM 7090 (whose computational power was roughly 0.05 MFLOPS \cite{hockney1988parallel}, whereas a top-range laptop is now capable of 100 GFLOPS) for a 200-activity project was 20 minutes for the case where a triangular distribution is used to model activity durations and 5 minutes for the uniform case \cite{vanSlyke} (the impact of the right choice of the probability distribution of activities is investigated in \cite{kirytopoulos2008PERT}). 

\section{A comparison for practical cases}
\label{sec:pratica}
The cases considered for a numerical analysis in Sections \ref{sec:interval} and \ref{sec:point} concern projects with some hundred or slightly more than one thousand activities. Since the computation time needed to arrive at an estimate of the project completion time depends on the project size, it is important to consider realistic size. 

\begin{figure}[htbp]
\centering
\begin{tikzpicture}
\begin{axis}[
cycle list name=mylist,
xmin=0,
xmax=500,
domain=0:500,
xlabel={Number of workpackages},
ylabel={Probability Density Function},
]
\addplot [black, mark=none] table[col sep=comma] {20170421-Nwpden.csv};
\end{axis}
\end{tikzpicture}
\caption{Empirical probability density function of the number of workpackages in a  project}
\label{fig:wpden}
\end{figure}
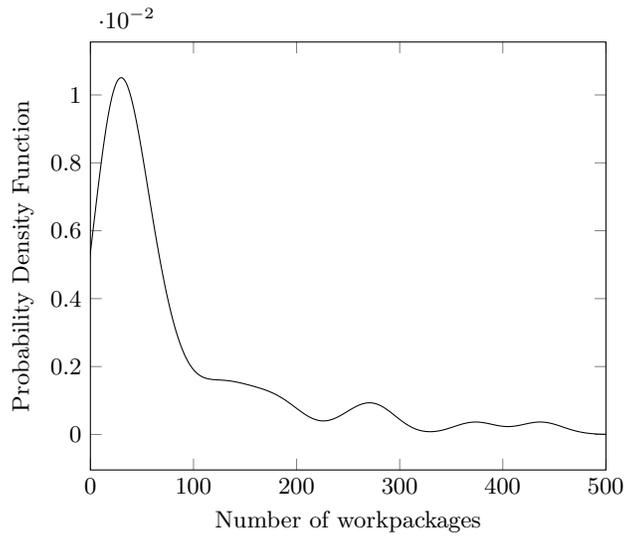

A host of data concerning real projects (baseline scheduling data, risk analysis data, and project control data) is available on the website \url{http://www.projectmanagement.ugent.be/?q=research/data/realdata}, which allow us to understand what figures we should use to carry out a realistic assessment. The project database hosted on that website is described in \cite{vanhoucke2016overview,batselier2015construction}. We extracted data concerning 52 projects from that database. The database does not include a graph representation of the projects involved, but provides the Work Breakdown Structure as a hierarchical list of Workpackages. We have assumed a one-to-one correspondence between workpackages and activities. In Table \ref{tab:statwp} we report the main statistics concerning the number of workpackages, which we therefore consider as statistics on the number of activities as well.
In \figurename~\ref{fig:wpden} we have plotted the empirical probability density function of the number of workpackages (hence of activities), obtained through the Gaussian kernel method \cite{silverman1986density}.

\begin{table}[]
    \centering
    \begin{tabular}{ccccc}
    \toprule
       Min & Mode & Mean   & Max & Std Dev \\
       \midrule
         7 & 29.58 & 80.58  & 437 & 94.81\\
         \bottomrule\\
    \end{tabular}
    \caption{Statistics of the number of workpackages in a project}
    \label{tab:statwp}
\end{table}

The values are well within the range considered in the studies mentioned so far: figures in the order of some hundred well represent the size of a typical project.

\section{Conclusions}
For the typical number of activities found in projects, the computational time to compute bounds does not represent a real problem. It may require a larger amount of time if we opt for a full MonteCarlo simulation, which provides us with a more complete characterization of the project makespan than those represented by the couple of bounds. However, embracing the MonteCarlo approach makes it easier to adopt a more complex (and realistic) model of the project, including, e.g., a specific correlation model between activities or more complex constraints. Though that could mean a further increase in the computation time, an accelerated MonteCarlo simulation could improve the performance and still keep the computation time within an acceptable range. 

The relevant research task should therefore concentrate on the search for a simpler estimation method, which can be easily handled by practitioners. Under this perspective, the  estimation time can be traded off for the simplicity. The key point is how to arrive at a simpler description of the project characteristics (activities durations and correlation) without sacrificing the accuracy of the estimate.












\end{document}